\documentclass[aps,prb,superscriptaddress,showpacs,floatfix]{revtex4-1}
\usepackage{graphicx,amsmath,amssymb,xspace,epsfig,float,multirow,subfigure,tabularx}
\usepackage{amsbsy,amssymb,amsmath,bm}
\usepackage{epsf}
\usepackage{color,natbib}
\usepackage{amsfonts}
\usepackage{hyperref}

\pdfoutput=1

\begin{document}

\title{Origin of the maximal critical temperature disparities in one - layer
cuprate superconductors.}
\author{Baruch Rosenstein}
\affiliation{Electrophysics Department, National Yang Ming Chiao Tung 
University, Hsinchu 30050, Taiwan, R. O. C}

\email{vortexbar@yahoo.com}
\author{B. Ya. Shapiro}
\affiliation{Physics Department, Bar-Ilan University, 52900 Ramat-Gan,
Israel}
\email{shapib@biu.ac.il}

\begin{abstract}
Recently a phonon exchange d - wave pairing mechanism in cuprates was
proposed. The phonons are the lateral apical oxygen atoms vibrations. They
generate the attractive pairing potential peaked at $\Gamma $ point of the
Brillouin zone, $V\left( k\right) \propto \exp \left[ -2kd_{a}\right] $,
where $d_{a}$ is distance from the $CuO$ planes. The model explains a rather
paradoxical well known negative correlation of\ the optimal doping critical
temperature $T_{c}^{\max }$ with $d_{a}$ and increase of $T_{c}^{\max }$
with pressure. However the large disparities in $T_{c}^{\max }$, especially
in one - layer cuprate superconductors, from $39K$ for $%
La_{2-x}Sr_{x}CuO_{4} $, to $95K$ for $HgBa_{2}CuO_{4+\delta }$, cannot be
attributed by differencies in $d_{a}$. Other important material parameters
include the hopping amplitudes $t,t^{\prime }$ and the on site Coulomb
repulsion $U$. It is shown (within weak coupling).that $T_{c}^{\max }$ is
highest for materials close to the topological (Lifshitz) transition from an
open to a close Fermi surface. The transition occurss for $t^{\prime
}=-0.19t $ and rather small values of effective value of $U=2t$ at optimal
doping. Analytic expressions for $T_{c}^{\max }$ are derived both near
criticality and away from it.
\end{abstract}

\pacs{PACS: 74.20.Rp,   74.72.-h,  74.25.Dw}
\maketitle

\affiliation{Electrophysics Department, National Yang Ming Chiao Tung 
University, Hsinchu 30050, Taiwan, R. O. C}

\affiliation{Physics Department, Bar-Ilan University, 52900 Ramat-Gan,
Israel}

\affiliation{Electrophysics Department, National Yang Ming Chiao Tung 
University, Hsinchu 30050, Taiwan, R. O. C}

\affiliation{Physics Department, Bar-Ilan University, 52900 Ramat-Gan,
Israel}

%\keywords{superconductivity, two dimensional electron gas, optical phonons,
%FeSe on STO}

\section{Introduction}

The physical cause of high temperature superconductivity in cuprates is
still under intensive debate despite enormous body of experimental facts and
numerous theoretical proposals. It is generally accepted that the normal
state including the Mott anti - ferromagnetic (AF) insulator at low doping,
the pseudogap and strange metal phases at intermediate dopings are due to
strong on site electron repulsion $U$ in the $CuO_{2}$ planes. Since the
pairing is d-wave, it was conjectured that an in - plane mechanism like the
AF spin fluctuations play a major role. One of the obvious questions one
asks about the pairing mechanism in cuprates\cite{Kaminski08} is why
structurally identical members of the cuprate family like the one - layered $%
La_{2-x}Sr_{x}CuO_{4}$ ("La") and $HgBa_{2}CuO_{4+\delta }$ ("Hg") have so
different maximal critical temperatures ($39K$ and $95K$ respectively). A
purely $CuO$ mechanism when applied to whole family of cuprates raises
several questions. Moreover it is believed now that La has \textit{larger} $%
U $ then Hg, so that a weaker coupled system develops much larger pairing
then an "iconic" strongly coupled system La.

Recent first principle calculations provide increasingly accurate estimates
of $U$ that should be used in Hubbard - like modeling on the "mesoscopic"
scale. Limiting ourselves to simplest one - layer cuprates, one estimates%
\cite{DFTsmallU} $U=6.5t$ for lower $T_{c}$ parent compounds $La_{2}CuO_{4}$
and $Bi_{2}Sr_{2}CuO_{6}$, while the highest $T_{c}^{\max }$ compounds, Hg
and $Tl_{2}Ba_{2}CuO_{6}$ have smaller on site Coulomb repulsion $U=4.5t$,
see Table 1. The nearest neighbors hoping is roughly the same ( $%
t=450-470meV $) for all the materials. Other recent calculations\cite%
{DFTlargeU} give slightly larger values of $U$, but ratios of $U$ for
various materials are the same. The critical temperature is maximal at the
universal hole doping, $p_{opt}\simeq 0.16$, for the four compounds. In the
framework of "unconventional" pairing mechanism (as the spin fluctuations
within the $CuO$ layer) the disparity in critical temperature is counter -
intuitive: stronger correlations should lead to larger $T_{c}^{\max }$ at
least in the intermediate coupling (namely $U<W$, where the bandwidth is $%
W\simeq 5-8t$). At very large $U$ the dependence of $T_{c}^{\max }$ is not
clear. Other presumably important characteristics of these materials in
addition to $t$ and $U$ are the apical distance and the next to nearest
hopping amplitudes determining the dispersion relation.

The structure of the compounds is quite similar. The lattice spacing is
almost the same, $a=3.8-3.9A$, while the two apical oxygen atoms are above
and below the $Cu$ atoms. These distances are $d_{a}=2.4A$ for the low $%
T_{c} $ materials (La, Bi) and$\ d_{a}=2.8A$ for higher $T_{c}$ (Hg,Tl), see
Table 1. Correlation of $T_{c}^{\max }$ with apical distance $d_{a}$
(despite the small differences between materials) and the next to nearest
neighbor hoping, $t^{\prime }$, was noticed early on in numerous cuprate
families\cite{Pavarini01}. On the basis of ARPES measurement it was noted%
\cite{Kaminski08}, that near the optimal doping the Fermi surfaces have
different topology: the diamond shape ("closed") for La and Bi, while it is
hyperbolically ("open") for $Hg$ and $Tl$, see Fig.1. This fact was first to
blame for a factor $2.5$ disparity in critical temperature. To quote\cite%
{Kaminski08}: "Since the size of the superconducting gap is largest in the
antinodal region, differences in the band dispersion at the antinode may
play a significant role in the pairing and therefore affect the maximum
transition temperature". This is also qualitatively in accord with values of 
$t^{\prime }$ estimated\cite{Pavarini01}, $\left \vert t^{\prime }\right
\vert /t=0.17$ in La and $0.33$ in Hg since at certain value of $t^{\prime
}/t$ there is a topological (Lifshitz) transition from an open to close
Fermi surface, see Fig.1, where the FS for $p_{opt}=0.16$ is given for five
values of \ $t^{\prime }$. It is not clear how close the two materials are
to the topological transition point (center panel in Fig.1), but apparently
La,Bi are far from it (on the left), while Tl,Hg are much closer (the right)
to the transition.

\begin{figure}[h]
\centering \includegraphics[width=18cm]{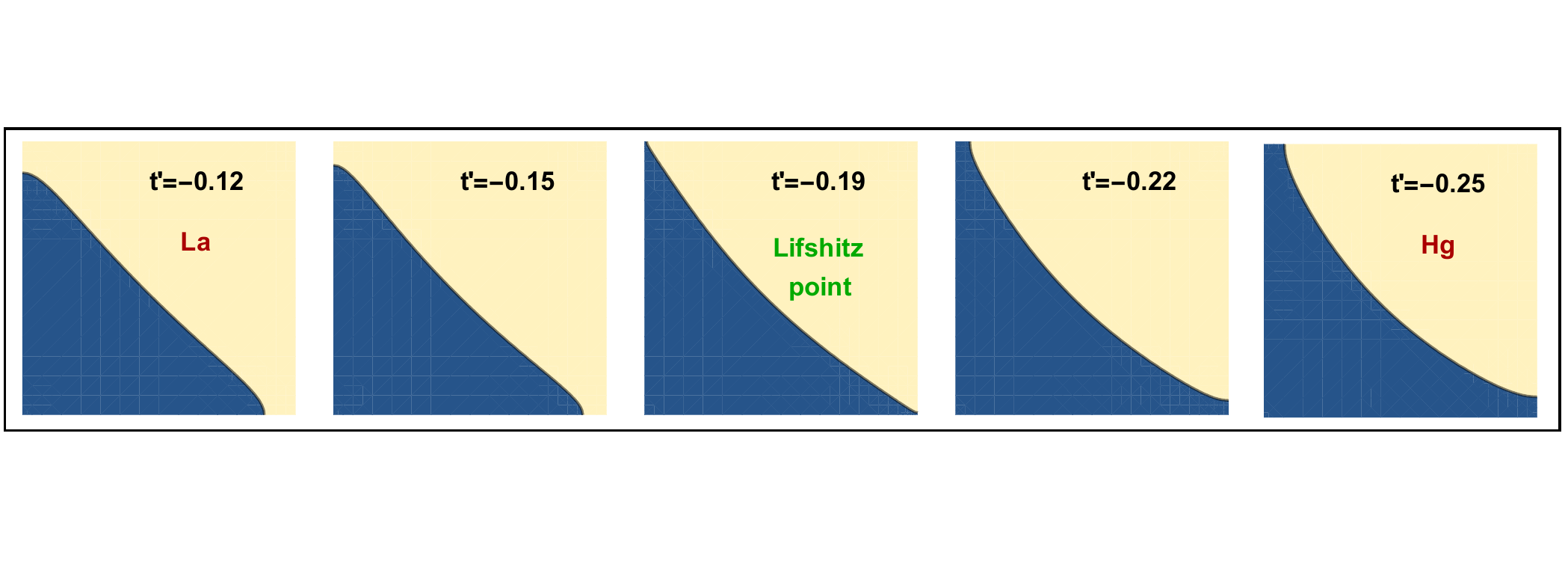}
\caption{ Fermi surface (in the first quarter of the BZ) of the tight
binding binding model at fixed doping (here the universal optimal doping
p=0.16). Topological transition occurs on the central frame. FL is closed
around the $\Gamma $ point on the left, while open (equivalently centered at
the $M$ point) on the right.}
\end{figure}

An idea that the peak in $T_{c}$ appears on the topological transition line
is not unique to the high $T_{c}$ cuprates. For example it was predicted\cite%
{WSM17} that when a Weyl semimetal undergoes a Lifshitz transition from the
type I (closed FS, elliptic) to the type II (open FS, hyperbolic) such a
peak (the s-wave) should appear. It was observed in $HfTe_{5}$ by applying
pressure\cite{HfTe}. The main reason is clearly the spectral weight
enhancement at points at which the pairing occurs. Returning to cuprates,
one notices that at the open - close transition FS passes the van Hove point 
$\mathbf{X}=\left( \pi ,0\right) $ of the BZ, see central frame in Fig.1,
The van Hove singularity in cuprates was observed experimentally\cite%
{Fischer11}. Theoretically the van Hove singularity's effect on the s - wave
superconductivity was studied \cite{vHove} (in some cases in relation to
cuprates\cite{Tsuei}), but the result was that the enhancement was found to
be negligible.

The question of the superconductivity dependence on the dispersion relation
(like tuning $t^{\prime }$ to the van Hove singularity) within the spin
fluctuation pairing theory typically based on the Hubbard model was not
addressed for a number of reasons. First it has not been convincingly
established that the model has the d - wave ground state around $p_{opt}$ .
The weak coupling, $U<4t$, just leads to an exponentially small d - wave
instability\cite{Kivelson}, while recently it was demonstrated that at
strong coupling, $U>5t$, competing correlated states take over\cite{nodwave}%
). Second, the strongly coupled theory requires application of very
complicated methods\cite{Katzenelson} that makes a delicate fine tuning of
the band parameters nearly impossible.

Recently a phonon exchange d - wave pairing mechanism in cuprates was
proposed\cite{I}. It was argued that the only phonon modes capable of
generating the d -wave pairing in the $CuO$ plane are the lateral apical
oxygen atoms vibrations. Better studied the in - plane breathing and
buckling oxygen vibrations\cite{Bulut} and the apical oxygen c - direction
modes\cite{apicz} are not able to generate a significant d-wave $T_{c}$. The
apical lateral oxygen vibrations in ionic environment generate\cite{I} an
attractive pairing in - plane potential, $V\left( k\right) \propto \exp %
\left[ -2kd_{a}\right] $. This phonon exchange model will be referred to as
ALLP. The $d_{a}$ dependence explains a well known correlation of\ the
optimal doping critical temperature $T_{c}^{\max }$ with pressure,but, as
mentioned above is impossible to explain the large disparities in $%
T_{c}^{\max }$. \ 

The model at intermediate $U$ describes well\cite{I} various normal
properties in the doping range not too close to the Mott insulator. In an
underdoped cuprate (with locally anti - ferromagnetic pseudogap) FS have
four Fermi pockets, while the transition to the overdoped samples occurs via
the Lifshitz point \ (optimal doping, \textit{different} from the open -
close FS Lifshitz transition discussed above) where the pockets coalesce
into a single Fermi surface. Importantly the phonon generated $T_{c}$ is
maximal at the optimal doping, where the pseudogap disappears. In the
overdoped strange metal linear in temperature resistivity was obtained\cite%
{preprint}.

In the present paper we apply the ALLP of the d - wave superconductivity at
optimal doping in the case when the system is near the open-close FS
topological transition. The main feature in this case is that of the Fermi
surface approaches the van Hove singularity point $\mathbf{X}$. In
particular it paradoxically results in the largest $T_{c}^{\max }$ for
materials with lowest $U$. The paper is organized as follows. In Section II
a model of the lateral optical phonons in ionic crystal and an effective $%
t-t^{\prime }$ model of the correlated electron gas is presented. This $%
50mev $ phonon mode and its coupling including the matrix elements are
described sufficiently well by the Born - Meyer approximation that has been
applied to cuprates\cite{Falter}. In Section III symmetrized Hartree - Fock
approximation\cite{Li19} (valid for intermediate strength $U$) is used to
investigate the Fermi surface topology change. We concentrate on the optimal
doping and consider the topological close to open Fermi surface transition.
In Section IV we develop a BCS - like d - wave theory of superconductivity
in the framework of dynamic approach. The effect of the van Hove singularity
is taken into account approximately. Section V contains results for
different one - layer cuprates obtained directly from numerical solution of
Gorkov equations. In the last Section results are summarized and discussed.

\section{The model}

Our model consists of the 2DEG interacting with phonons of a polar insulator:%
\begin{equation}
H=H_{e}+H_{ph}+H_{e-ph}\text{.}  \label{Hamiltoniandef}
\end{equation}%
We start with the phonon term. The electron part is the Hubbard model, while
the coupling between the electronic and vibrational degrees of freedom, $%
H_{e-ph}$, is subject of the last Subsection. 
\begin{table*}[h]
\caption{Material parameters of one layer cuprates.}
\begin{center}
\begin{tabular}{|l|l|l|l|l|}
\hline
material & $La_{2-x}Sr_{x}CuO_{4}$ & $Bi_{2}Sr_{2}CuO_{6+\delta }$ & $%
HgBa_{2}CuO_{4+\delta }$ & $Tl_{2}Ba_{2}CuO_{6+\delta }$ \\ \hline
$T_{c}^{opt}$ & $39$ & $35$ & $95$ & $95$ \\ \hline
$U_{parent}/t$ & $6.5$ & $4.06$ & $3.93$ & $3.82$ \\ \hline
$\overline{U}_{opt}/t$ & $2.6$ & $2.7$ & $1.9$ & $1.9$ \\ \hline
$t^{\prime }/t$ & $-0.13$ & $-0.12$ & $-0.19$ & $-0.20$ \\ \hline
$d_{a}\left[ A\right] $ & $2.4$ & $2.45$ & $2.79$ & $2.7$ \\ \hline
\end{tabular}%
\end{center}
\end{table*}

\subsection{The apical oxygen lateral vibrations contribution to the
effective electron - electron interaction}

Although the prevailing hypothesis is that superconductivity in cuprate is
"unconventional", namely not to be phonon - mediated, the phonon based
mechanism has always been a natural option to explain extraordinary
superconductivity in cuprates. The crystal has very rich spectrum of phonon
modes. However very few have a strong coupling to 2DEG and even fewer can
generate lateral (in plane) forces causing pairing. As mentioned in
Introduction, the most studied phonon "glue" mode has been the oxygen
vibrations within the $CuO$ plane\cite{Bulut}. It was argued\cite{Rosen19}
(in the context of high $T_{c}$ 1UC $FeSe$ on perovskite substrates where
this was observed experimentally\cite{Guo}) that lateral vibrations of the
oxygen atoms in the adjacent ionic perovskite layer can couple strongly to
2DEG residing in the $CuO$ plane.

The structure of the $La_{2}CuO_{4}$ and $HgBa_{2}CuO_{6}$ unit cell below
the conducting layer is schematically depicted in Fig. 2. Besides the single 
$CuO_{2}$ ($Cu$ is drawn in Fig.2 as a brown sphere, $O$ - small orange
spheres).layer only two insulating oxide layers are assumed to be relevant.
The closest layer at distance $d_{a}$, see Table I, consists of heavy $Ba$
atoms (cyan rings) and light "apical" oxygen (small red circle). The next
layer is $LaO/HgO$, ($Hg$ - violet, $O$ - small dark red circles).
Qualitatively one of the reasons is that the $LaO/BaO$ layer, see Fig. 2,
constitutes a strongly coupled ionic insulator. Unlike the metallic layer
where screening is strong, in an ionic layer screening is practically absent
and a simple microscopic theory of phonons and their coupling exists\cite%
{Abrahamson}. Although phonons in cuprates were extensively studied within
the first principle approach including the oxygen vibration mode the matrix
elements are rarely discussed, so that the simple approach is appropriate%
\cite{Falter}. It was repeatedly noticed\cite{Gorkov} that vibrations in $c$
directions contribute little to the in - plane pairing.

\begin{figure}[h]
\centering \includegraphics[width=16cm]{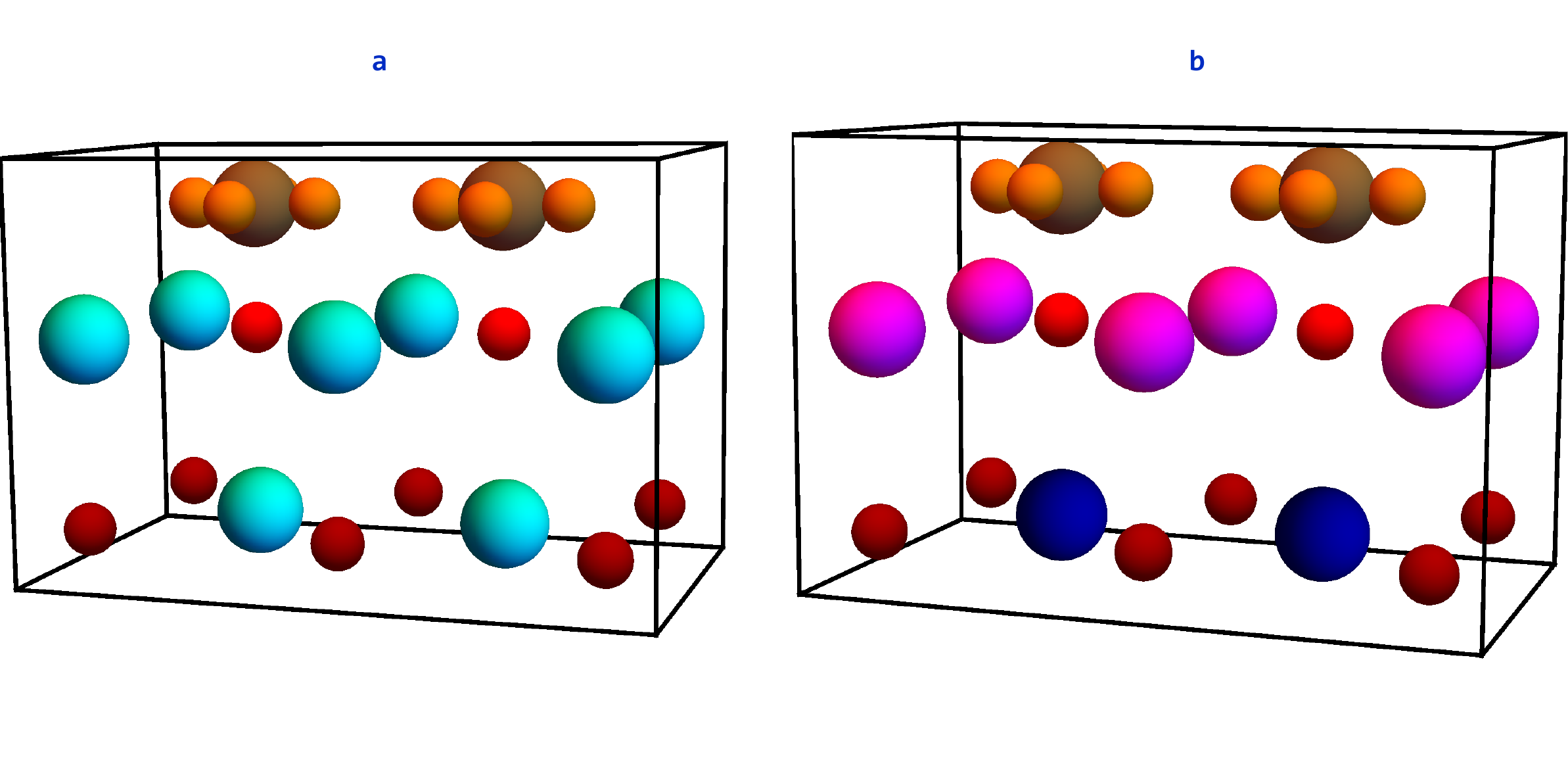}
\caption{ The profile 3D view of three layers comprising relevant part of
the one unit cell :of $La_{2}CuO_{4}$ and $HgBa_{2}CuO_{6}$. The top is the $%
CuO_{2}$ ($Cu$ - brown sphere, $O$ - orange), ALLP are located inthe
insulating $LaO/BaO$ ($La$ -cyan $Ba$ - magenta violet) $O$ (dark red). The
third layer: $LaO/HgO$, ($Hg$ - violet, $O$ - red). Sizes of atoms are
inversely proportional to the values of the Born - Mayer inter - atomic
potential parameter parameter $b$ }
\end{figure}

Phonons in ionic crystals are described sufficiently well by the Born -
Meyer potential due to electron's shells repulsion\cite{Abrahamson} and
electrostatic interaction of ionic charge,%
\begin{equation}
V\left( r\right) =\sqrt{A_{1}A_{2}}\exp \left[ -\frac{1}{2}\left(
b^{1}+b^{2}\right) r\right] +Z_{1}Z_{2}\frac{e^{2}}{r}\text{,}
\label{interatomic}
\end{equation}%
with standard values of coefficients $A$ and $b$ for all the atoms. The
ionic charges $Z$ are estimated from the DFT calculated Milliken charges\cite%
{averestov} or determined phenomenologically. In the $LaO/BaO$ layer the
charges are constrained by neutrality. Since oxygen is much lighter than $%
La/Ba$, the heavy atoms' vibrations are negligible. Obviously that way we
lose the acoustic branch, however it is known that the acoustic phonons
contribute little to the pairing\cite{Gorkov}. Atoms in neighboring layers
can also be treated as static. Moreover one can neglect more distant layers.
Even the influence of the yet lower layers (below the last layer shown in
Fig. 2) is insignificant due to the distance. Consequently the dominant
lateral displacements, $u_{\mathbf{m}}^{\alpha }$, $\alpha =x,y$, are of the
oxygen atoms directly beneath the $Cu$ sites at $\mathbf{r}_{\mathbf{m}%
}=a\left( m_{1},m_{2}\right) $.

The dynamic matrix $D_{\mathbf{q}}^{\alpha \beta }$ is calculated by
expansion of the energy to second order in oxygen displacement (details in
Appendix A in ref.\cite{I}), so that the phonon Hamiltonian in harmonic
approximation is:

\begin{equation}
H_{ph}=\frac{1}{2}\sum \nolimits_{\mathbf{q}}\left \{ M\frac{du_{-\mathbf{q}%
}^{\alpha }}{dt}\frac{du_{\mathbf{q}}^{\alpha }}{dt}+u_{\mathbf{-q}}^{\alpha
}D_{\mathbf{q}}^{\alpha \beta }u_{\mathbf{q}}^{\beta }\right \} \text{.}
\label{Hph}
\end{equation}%
Here $M$ is the oxygen mass. Summations over repeated components indices is
implied. Now we turn to derivation of the phonon spectrum. Two eigenvalues,
the transversal optical (TO) and the longitudinal optical (LO) modes are
given in obtained. One observes that there are longitudinal modes are in the
range $\Omega _{\mathbf{q}}\sim $ $45-511mev$ and $25-32mev$ respectively.
The energy of LO modes is larger than that of the corresponding TO, although
the sum $\Omega _{\mathbf{q}}^{LO}+\Omega _{\mathbf{q}}^{TO}$ is nearly
dispersionless. At $\Gamma $ the splitting is small, while due to the long
range Coulomb interaction there is hardening of LO and softening of TO at
the BZ edges. The dispersion of the high frequency modes is small, while for
the lower frequency mode it is more pronounced.

\subsection{The $t-t^{\prime }$ Hubbard model of the 2DEG in $CuO$ layers.}

The electron gas is described by an effective single band $t-t^{\prime }$
model. Hamiltonian in momentum space is:

\begin{equation}
H_{e}=\sum \nolimits_{\mathbf{k}}c_{\mathbf{k}}^{\sigma \dagger }\left(
\epsilon _{\mathbf{k}}-\mu \right) c_{\mathbf{k}}^{\sigma }+U\sum \nolimits_{%
\mathbf{i}}n_{\mathbf{i}}^{\uparrow }n_{\mathbf{i}}^{\downarrow }\text{,}
\label{tight1}
\end{equation}%
Here $c_{\mathbf{k}}^{\sigma \dagger }$ is the electron creation operator
with spin projection $\sigma =\uparrow ,\downarrow $. Only nearest and next
to nearest neighbors hopping terms are included: 
\begin{equation}
\epsilon _{\mathbf{k}}=-2t\left( \cos \left[ ak_{x}\right] +\cos \left[
ak_{y}\right] \right) -4t^{\prime }\cos \left[ ak_{x}\right] \cos \left[
ak_{y}\right] \text{.}  \label{eps}
\end{equation}%
The dispersion relation thus is simplified with respect to a "realistic" one
in which more distant hops are included. The on site repulsion is described
by the on site Hubbard repulsion term with $n_{\mathbf{i}}^{\sigma }=c_{%
\mathbf{i}}^{\sigma \dagger }c_{\mathbf{i}}^{\sigma }$ being the spin $%
\sigma $ occupation on the site $\left \{ i_{x},i_{y}\right \} $ and $\mu $
is the chemical potential. Due to the repulsion, even the model without
phonons is nontrivial and will be treated approximately in the next Section.
Now we turn to the electron - phonon coupling.

\subsection{Electron - phonon coupling}

The lateral apical oxygen phonon's interaction with the 2DEG on the adjacent 
$CuO$ layer $d_{a}$ above the $LaO/BaO$ plane is determined by the electric
potential created the charged apical oxygen vibration mode $\mathbf{u}_{%
\mathbf{m}}$ at arbitrary point $\mathbf{r}$ is:

\begin{equation}
\Phi \left( \mathbf{r}\right) =\sum \nolimits_{\mathbf{m}}\frac{Ze}{\sqrt{%
\left( \mathbf{r}-\mathbf{r}_{\mathbf{m}}-\mathbf{u}_{\mathbf{m}}\right)
^{2}+d_{a}^{2}}}\text{,}  \label{3_pot}
\end{equation}%
The apical oxygen charge was taken as $Z=1.3$. The electron-phonon\
interaction (EPI) Hamiltonian that accounts for the hole charge distribution
in the $CuO$ plane is derived in Appendix A of ref.\cite{I}. The result in
momentum space has a density - displacement form

\begin{equation}
H_{eph}=Ze^{2}\sum \nolimits_{\mathbf{q}}n_{-\mathbf{q}}g_{\mathbf{q}%
}^{\alpha }u_{\mathbf{q}}^{A\alpha }\text{,}  \label{Heph}
\end{equation}%
with EPI matrix element, 
\begin{equation}
\mathbf{g}_{\mathbf{q}}=2\pi \widehat{\mathbf{q}}e^{-\left \vert q\right
\vert d_{a}}\text{.}  \label{A}
\end{equation}%
Here $\left \vert \mathbf{q}\right \vert ^{2}\equiv 4\left( \sin ^{2}\left[
q_{x}/2\right] +\sin ^{2}\left[ q_{y}/2\right] \right) $. It is well known
that only longitudinal phonons contribute to the effective electron -
electron interaction, as is clear from the scalar product form of the Eq.(%
\ref{Heph}). The precision of the last equality is 2\%, see figure 9,10 in
Appendix A of ref.\cite{I}.

To conclude Eqs.(\ref{tight1},\ref{Hph},\ref{Heph}) define our microscopic
model. Now we turn to description of the normal state

\section{Tuning to the open - close Fermi surface transition}

\subsection{ Mean field description of the Hubbard model at intermediate
coupling}

The normal state of cuprates exhibits a host of phenomena including
pseudogap in underdoped regime resulting in fracture of the Fermi surface.
In the intermediate to weak coupling range, $U/t=1-2.5$, one can use a much
simpler approximation scheme: the Hartree - Fock approximation with its
extension including the RPA type coupling $U$ renormalization \cite%
{Maier07,I} and symmetrized HF\cite{Li19}. It provides a good agreement with
the more sophisticated methods. The coupling $U$ that enters the Hubbard
description generally depends on doping. It is largest for the parent
material (Mott insulator in all holed doped cuprates). The first principle
values were given in Introduction. However in the present paper we focus on
the optimal doping $p_{opt}=0.16$ for which the value $\overline{U}$ is
significantly reduced. In some first principle calculations doping is
simulated by periodic substitution. For example the band structure of the $%
p=0.25$ La was simulated recently\cite{Markiewcz18}. As discussed above host
of normal and superconducting properties in overdoped, optimal and slightly
underdoped materials can be described in the intermediate coupling regime.

Generally the HF approximation takes into account the ground state
reconstruction due to the electron - hole pairs generated by the Coulomb
repulsion. In translation invariant case it results in renormalization of
the chemical potential%
\begin{equation}
\mu _{r}=\mu -Un/2\text{.}  \label{muren}
\end{equation}%
Within the HF approximation the density at temperature $T$ at optimal (and
above) doping is related to the dispersion relation and chemical potential $%
\mu $\ by%
\begin{equation}
n=\frac{2}{\left( 2\pi \right) ^{2}}\int_{BZ}\frac{1}{1+\exp \left[ \left(
\epsilon _{\mathbf{k}}-\mu _{r}\right) /T\right] }\text{.}  \label{densityHF}
\end{equation}%
From now on we chose units in which $a=t=\hbar =1$.

Let us find the value of parameters at which the renormalized FS "reaches"
the van Hove singularity point $\mathbf{X}=\left( \pi ,0\right) $, $\epsilon
_{\mathbf{X}}-\mu _{r}=0$. As can be seen from Fig.3c and Fig.1, precisely
at this value the electronic system undergoes a Lifshitz transition from
closed FS (like in La) to an open one (like in Hg). For our dispersion
relation, Eq.(\ref{eps}), this becomes%
\begin{equation}
\mu _{r}=\epsilon _{\mathbf{X}}=4t^{\prime }\text{.}  \label{vHcond}
\end{equation}%
At this point Eq.(\ref{densityHF}) depends on $t^{\prime }$ only:%
\begin{equation}
n=\frac{2}{\left( 2\pi \right) ^{2}}\int_{BZ}\frac{1}{1+\exp \left[ \left(
\epsilon _{\mathbf{k}}-\epsilon _{\mathbf{X}}\right) /T\right] }.
\label{den}
\end{equation}%
The doping $p=1-n$ is given for $T=0.02$ $t$ in Fig.3. It is almost
independent of temperature for $T<0.1$ $t\approx 500K$ for $t=450$ $meV$ and
allows a linear fit, $p=0.84$ $t^{\prime }/t$. \ In particular for the
optimal doping $p=0.16$ one should have $t^{\prime }=-0.19$ $t$. If the Tl
and Hg cuprates are very close to the transition due their high $T_{c}^{\max
}$, they should have this value of the next to nearest neighbor hopping.

The assumption that the doping is optimal means that it obeys an additional
equation for the AF-paramagnet criticality (assuming the second order
transition), see ref. \cite{I}. This determines $U$ as the following
integral over the \textit{magnetic} BZ:%
\begin{equation}
U^{-1}=\frac{1}{4\left( 2\pi \right) ^{2}}\int_{k_{1}=0}^{\pi
}\int_{k_{y}=-\pi }^{\pi }\frac{1}{\left \vert h_{\mathbf{k}}\right \vert }%
\left( \tanh \left[ \frac{\varepsilon _{\mathbf{k}}^{\prime }-\mu _{r}+\left
\vert h_{\mathbf{k}}\right \vert }{2T}\right] -\tanh \left[ \frac{%
\varepsilon _{\mathbf{k}}^{\prime }-\mu _{r}-\left \vert h_{\mathbf{k}%
}\right \vert }{2T}\right] \right)  \label{Ucond}
\end{equation}%
Here $k_{1}=k_{x}-k_{y}$ and,

\begin{eqnarray}
h_{\mathbf{k}} &=&1+e^{i\left( 2k_{1}\right) }+e^{ik_{y}}+e^{i\left(
2k_{1}-k_{y}\right) };  \label{h} \\
\varepsilon _{\mathbf{k}}^{\prime } &=&-4t^{\prime }\cos \left[ k_{1}\right]
\cos \left[ k_{1}-k_{y}\right] \text{.}  \notag
\end{eqnarray}%
At van Hove chemical potential it takes a form:%
\begin{equation}
U^{-1}=\frac{1}{4\left( 2\pi \right) ^{2}}\int_{k_{1}=0}^{\pi
}\int_{k_{y}=-\pi }^{\pi }\frac{1}{\left \vert h_{\mathbf{k}}\right \vert }%
\left( \tanh \left[ \frac{\varepsilon _{\mathbf{k}}^{\prime }-4t^{\prime
}+\left \vert h_{\mathbf{k}}\right \vert }{2T}\right] -\tanh \left[ \frac{%
\varepsilon _{\mathbf{k}}^{\prime }-4t^{\prime }-\left \vert h_{\mathbf{k}%
}\right \vert }{2T}\right] \right)  \label{UvH}
\end{equation}%
This gives the coupling at optimal doping for the van Hove material when $%
t^{\prime }=-0.19$, $U_{vH}=1.74$ $t$, see Fig.3b. Away from the van Hove
value of $t^{\prime }$ one obtains couplings shown in Fig.3b. For smaller
values of $t^{\prime }$ corresponding to closed FS one obtains larger $U$
but still within the range of perturbatively. improved HF theory\cite%
{preprint}.

The range of acceptable values of $t^{\prime }/t$ is rather limited. If one
chooses $\left \vert t^{\prime }\right \vert /t<0.1$, the Mott state at very
low doping does not appear\cite{Irkhin16}. At values larger than $%
\left
\vert t^{\prime }\right \vert /t>0.25$ the shape of the Fermi surface
in the underdoped regime is qualitatively different from the one observed by
ARPES. The value of $t^{\prime }=-0.19t$ for Hg and Tl is chosen to tune the
Lifshitz (topological) transition from the full Fermi surface to the
fractured one (four arcs) occurs at experimentally observed doping $%
x^{opt}=0.16$.

.

\begin{figure}[h]
\centering \includegraphics[width=16cm]{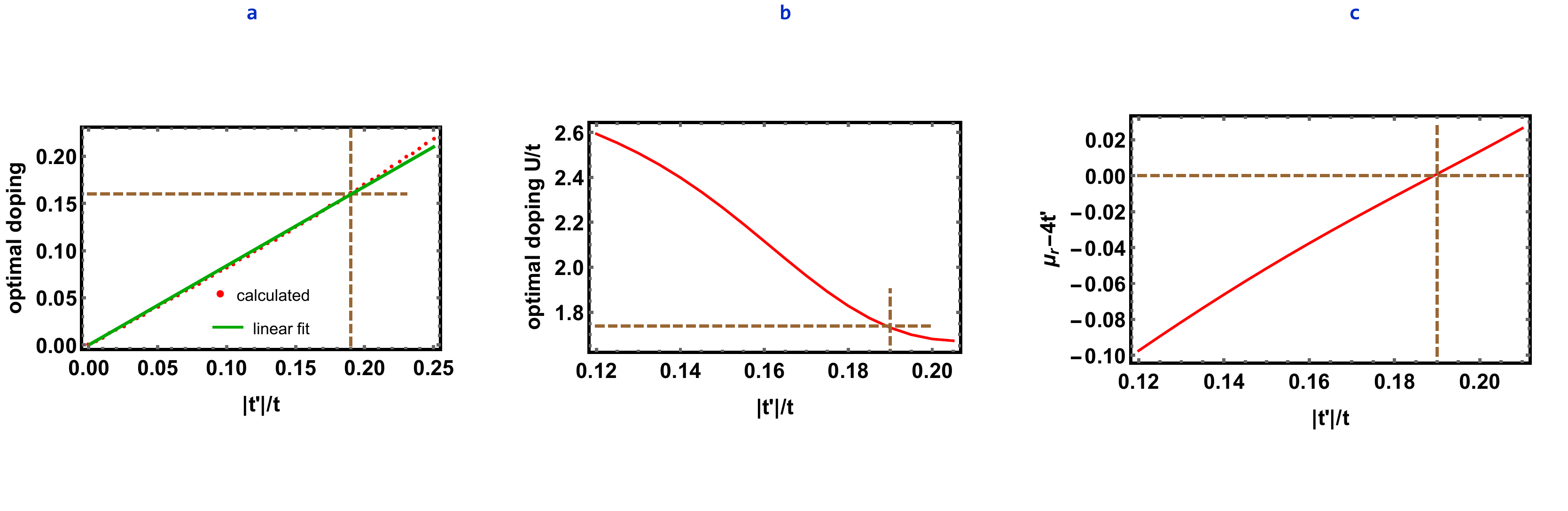}
\caption{a. The optimal doping \ as function of the ration of hoping
amplitudes $t^{\prime }/t$. b. An effective on site Coulomb repulsion. c.
Energy difference between the Femi level and the van Hove energy. .}
\end{figure}

\section{The d-wave phonon mediated superconductivity.}

It is well established that phonons cause s - wave pairing in low $T_{c}$
materials, however d-wave pairing is possible when forward scattering peak
of Eq.(\ref{A}) is present. Early work in this direction was summarized in
ref. \cite{Kulicrev}. The transition temperature $T_{c}$ as function of the
hole doping within the present model was studied in ref. \cite{I}
demonstrates the maximum at $p_{opt}$. We start from the derivation of the
phonon exchange d wave "potential" (appearing mainly near the $\Gamma $
point of BZ).

\begin{figure}[h]
\centering \includegraphics[width=18cm]{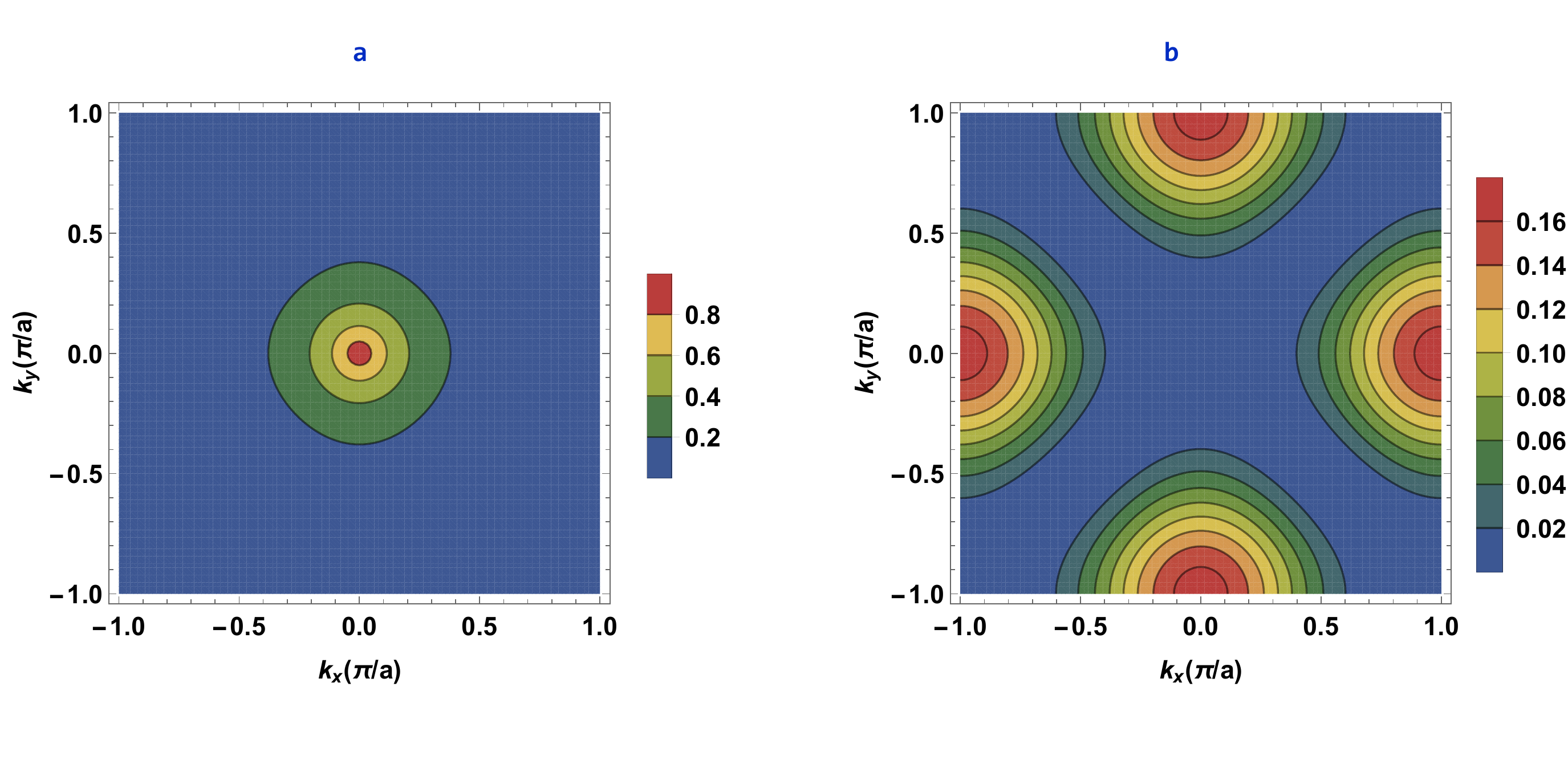}
\caption{a. Effective electron - electron attraction potential. Maximum at $%
\mathbf{q=0}$ is the result of electrostatics due to the distance $d_{a}$.
point. b. The effective potential for the d - wave pairing corresponding to
the ALLP potential. The peak now moved to points $\mathbf{X}=\left( \protect%
\pi /a,0\right) $ and $\mathbf{Y}=\left( 0,\protect \pi /a\right) $}
\label{Fig,6}
\end{figure}

\subsection{Effective phonon generated electron - electron interactions in
spin singlet channel}

In order to describe superconductivity, one should "integrate out" the
phonon and the spin fluctuations degrees of freedom to calculate the
effective electron - electron interaction. We start with the phonons. The
Matsubara action for EPI, Eq.(\ref{Heph}), and phonons, Eq.(\ref{Hph}), are,%
\begin{equation}
\frac{1}{T}\sum \nolimits_{m\mathbf{,q}}\left( Ze^{2}n_{-m,-\mathbf{q}}g_{%
\mathbf{q}}^{\alpha }u_{m,\mathbf{q}}^{\alpha }+\frac{M}{2}u_{-m,-\mathbf{q}%
}^{\alpha }\Pi _{m,\mathbf{q}}^{\alpha \beta }u_{m\mathbf{,q}}^{\beta
}\right) \text{,}  \label{action}
\end{equation}%
where $n_{-n,-\mathbf{q}}=\sum \nolimits_{\mathbf{k,}m}\psi _{\mathbf{k}-%
\mathbf{q},m-n}^{\ast \sigma }\psi _{\mathbf{k},m}^{\sigma }$ and $\mathbf{g}
$ was defined in Eq.(\ref{A}). The polarization matrix is defined via the
dynamic matrix of Eq.(\ref{Hph}): $\Pi _{n,\mathbf{q}}^{\alpha \beta
}=\left( \omega _{n}^{b}\right) ^{2}\delta _{\alpha \beta }+M^{-1}D_{\mathbf{%
q}}^{\alpha \beta }$, $\alpha ,\beta =x,y$. Since the action is quadratic in
the phonon field $\mathbf{u}$, the partition function is gaussian and can be
integrated out exactly, see details in ref.\cite{Rosen19}. As a result one
obtains the effective density - density interaction term for of electrons

\begin{eqnarray}
\mathcal{A}_{eff}^{ph} &=&\frac{1}{2T}\sum \nolimits_{\mathbf{q},n}n_{n,%
\mathbf{q}}V_{n\mathbf{q}}n_{-n,-\mathbf{q}};\text{ \ }V_{n,\mathbf{q}}=-%
\frac{\gamma }{\omega _{n}^{b2}+\Omega ^{2}}e^{-2d_{a}\left \vert \mathbf{q}%
\right \vert }+U;  \label{Aee} \\
\text{\  \  \  \  \ }\gamma &=&\frac{\left( 2\pi Ze^{2}\right) ^{2}}{M}\text{.} 
\notag
\end{eqnarray}%
The central peak is clearly seem in Fig. 4a.

The standard superconducting gap equation is,

\begin{equation}
\Delta _{m\mathbf{k}}=-T\sum \nolimits_{n\mathbf{p}}\frac{V_{m-n,\mathbf{k-p}%
}\Delta _{n\mathbf{p}}}{\omega _{n}^{2}+\left( \epsilon _{\mathbf{p}}-\mu
_{r}\right) ^{2}+\left \vert \Delta _{n\mathbf{p}}\right \vert ^{2}}\text{.}
\label{gapeqpara}
\end{equation}%
Here the (Matsubara) gap function is related to the anomalous GF, $\left
\langle \psi _{m\mathbf{k}}^{\sigma }\psi _{n\mathbf{p}}^{\rho }\right
\rangle =\delta _{n+m}\delta _{\mathbf{k+p}}\varepsilon ^{\sigma \rho }F_{m%
\mathbf{k}}$ ($\varepsilon ^{\sigma \rho }$ - the antisymmetric tensor), by 
\begin{equation}
\Delta _{m\mathbf{k}}=T\sum \nolimits_{n\mathbf{p}}V_{m-n,\mathbf{k-p}}F_{n%
\mathbf{p}}\text{.}  \label{deltasupdef}
\end{equation}%
The gap equation was solved numerically by iteration by discretization the
BZ with $N=256$ and $64$ frequencies. It converges to the d - wave solution.
The absolute value of the Matsubara gap function has a maximum near the
crystallographic $\mathbf{X}$ point, $\left( 0,\pi \right) $. At critical
temperature the gap disappears. The values for various hoping amplitudes at
optimal doping are presented in Fig.5 (as blue points). It can be fitted
well by a Lorenzian (blue dashed curve):%
\begin{equation}
T_{c}^{\max }=\frac{T_{L}}{1+\alpha \left( t^{\prime }-t_{L}^{\prime
}\right) ^{2}}\text{,}  \tag{TL}
\end{equation}%
with maximal value of temperature at the open to close Lifshitz point, $%
T_{L}=92K$ achieved next to nearest hoping amplitude $t_{L}^{\prime
}/t=-0.19 $. Other parameters characterizing the electron - phonon
interactions, coupling $U$ and the electron gas were given in Section II.
The Coefficient $\alpha $ is different for $\left \vert t^{\prime }\right
\vert <\left \vert t_{L}^{\prime }\right \vert $, $\alpha _{<}=180,$ and $%
\left \vert t^{\prime }\right \vert >\left \vert t_{L}^{\prime }\right \vert 
$, with $\alpha _{>}=240$.

To understand qualitatively the results, we derive next approximate formulas
for $T_{c}$ in terms of the "effective" d - wave pairing strength $\lambda
_{d}$ analogous to the well known s - wave coupling $\lambda _{s}$. The case
of the Lifshitz transition requires a special treatment near the van Hove
point.

\begin{figure}[h]
\centering \includegraphics[width=10cm]{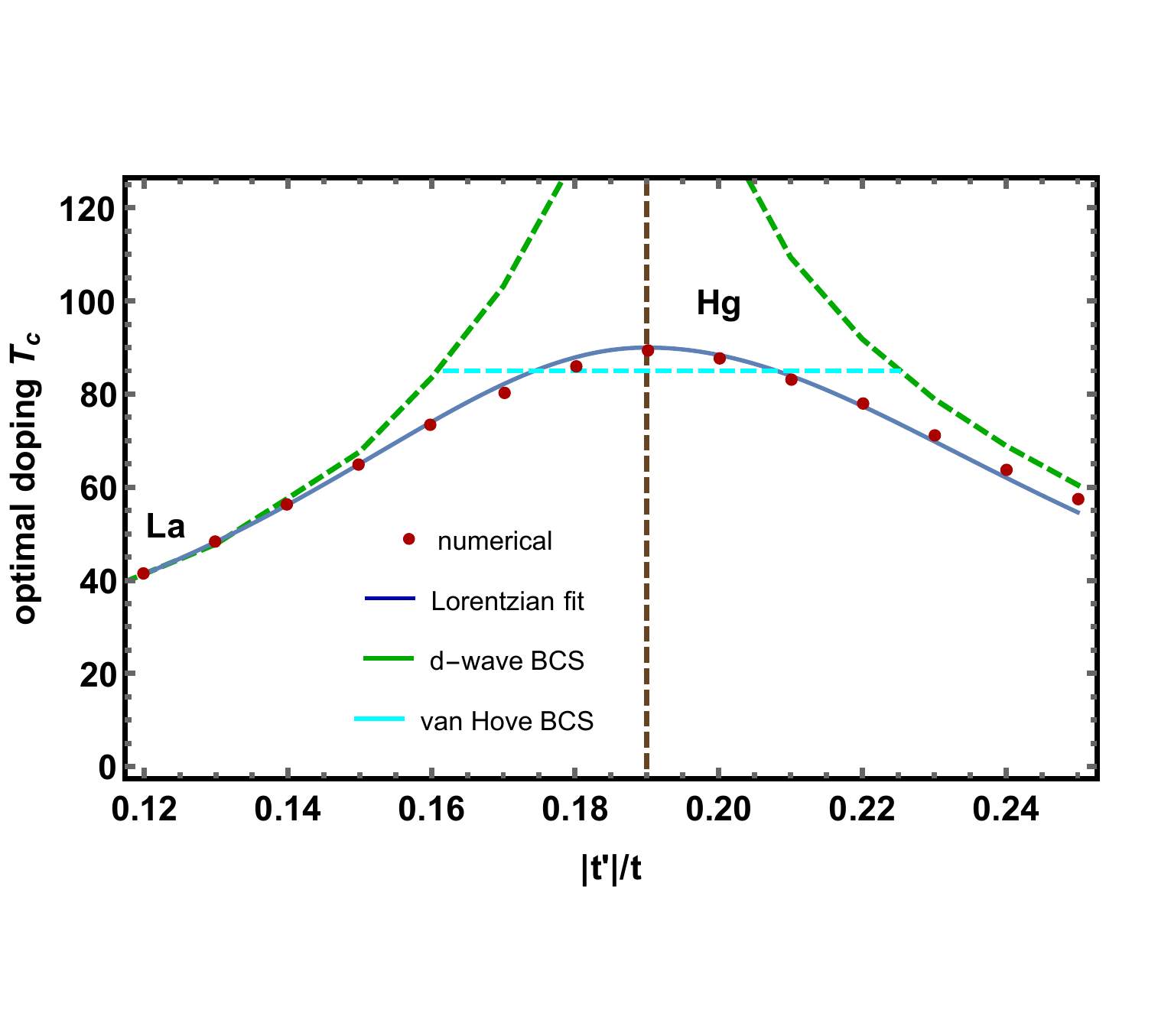}
\caption{Critical temperature at optimal doping as function of $t;/t$. \
Numerical solution is very close to the BCS formula (green dashed line) away
from topological criticality. The BCS expression is "cut off" by the van
Hove dominated critical value (light blue line). The numerical results are
fitted by Lorentzian (blue line) given in Eq.().}
\label{Fig,5}
\end{figure}

\subsection{The d -wave analog of the asymptotic expression for $T_{c}$}

Near $T_{c}$ one neglects the second term in denominator and obtains

\begin{equation}
\Delta _{m\mathbf{k}}=-T_{c}\sum \nolimits_{n\mathbf{p}}V_{m-n,\mathbf{k-p}}%
\frac{\  \Delta _{n\mathbf{p}}}{\omega _{n}^{2}+\left( \epsilon _{\mathbf{p}%
}-\mu _{r}\right) ^{2}}\text{.}  \label{Tcgap}
\end{equation}%
Numerical solution of the gap equation shows that the gap function decreases
slowly like $1/\omega $ at large $\omega $ and is often used at intermediate
coupling\cite{Maier} and at relevant relatively small frequencies can be
approximated by its $n=0$ component:%
\begin{equation}
\Delta _{n\mathbf{p}}=\Delta _{\mathbf{p}}=f_{d}\left( \mathbf{p}\right)
\Delta ,  \label{Maier}
\end{equation}%
The quasi - momentum dependence turns out to be quite universal:

\begin{equation}
f_{d}\left( \mathbf{p}\right) =\cos \left[ p_{x}\right] -\cos \left[ p_{y}%
\right] .  \label{fd}
\end{equation}%
This is an analog of the BSC assumption $\Delta _{n\mathbf{p}}=\Delta $ for
the s -wave. The normalization, $\sum \nolimits_{\mathbf{k}}f\left( \mathbf{k%
}\right) ^{2}=1$, is convenient for the following derivation.

After multiplication of the $T_{c}$ equation, Eq.(\ref{Tcgap}), by $f\left( 
\mathbf{p}\right) $ and summation over both $\mathbf{p}$ and $n$, it becomes:

\begin{equation}
1=-\sum \nolimits_{\mathbf{p}}\frac{\rho _{\mathbf{k-p}}^{d}}{\Omega ^{2}}\ 
\frac{\tanh \left[ \left( \epsilon _{\mathbf{p}}-\mu _{r}\right) /2T_{c}%
\right] }{2\left( \epsilon _{\mathbf{p}}-\mu _{r}\right) }\text{.}
\label{deltaeq}
\end{equation}%
Here 
\begin{equation}
\rho _{\mathbf{p}}^{d}=f_{d}\left( \mathbf{p}\right) \sum \nolimits_{\mathbf{%
k}}\rho _{\mathbf{k-p}}\ f_{d}\left( \mathbf{k}\right) ,  \label{rhod_def}
\end{equation}%
is an effective form factor" of the phonon mediated interaction, see Fig.4b.
For d-wave the nodal region is irrelevant, while the anti-nodal region is
strongly emphasized. For the s-wave it is simpler: $\rho _{\mathbf{p}}^{s}=$ 
$\rho _{\mathbf{p}}$.

Function $\tanh \left[ \varepsilon /2T_{c}\right] /\varepsilon $ is peaked
at $\varepsilon =0$ with width $T_{c}$. Using steepest descent approximation
(the function $\rho _{\mathbf{k}}^{d}$, drawn in Fig. 4b varies much
slower), one arrives at:

\begin{equation}
1\simeq -\frac{1}{2\left( 2\pi \right) ^{2}\Omega ^{2}}\int_{BZ}d^{2}\mathbf{%
p}\text{ }\rho _{\mathbf{p}}^{d}|_{\epsilon _{\mathbf{p}}=\mu _{r}}\frac{%
\tanh \left[ \left( \epsilon _{\mathbf{p}}-\mu _{r}\right) /2T_{c}\right] }{%
\epsilon _{\mathbf{p}}-\mu _{r}}\text{.}  \label{Tc1}
\end{equation}%
Making change of variables $\left \{ p_{x},p_{y}\right \} $ to geodesic
coordinates $\left \{ p_{n},p_{t}\right \} $ (the normal and the tangential
momenta), and then from $p_{n}$ to $\varepsilon =\epsilon \left(
p_{x},p_{y}\right) -\mu _{r}$, one obtains the BCS formula:

\begin{equation}
1=\lambda ^{d}\int_{\varepsilon =-\Omega }^{\Omega }d\varepsilon \frac{\tanh %
\left[ \varepsilon /2T_{c}\right] }{2\varepsilon }\simeq \lambda ^{d}\log 
\frac{2\Omega \gamma _{E}}{\pi T_{c}}\text{.}  \label{Tc2}
\end{equation}%
Here the d - wave coupling constant is defined by,%
\begin{equation}
\lambda ^{d}=-\frac{1}{\Omega ^{2}\left( 2\pi \right) ^{2}}%
\oint_{\varepsilon }dp_{t}\frac{1}{v\left( p_{t}\right) }\rho
_{p_{t},\varepsilon }^{d}=\frac{D\left( \mu _{r}\right) }{2\Omega ^{2}}\rho
_{p_{t},\varepsilon }^{d}\text{,}  \label{Tc3}
\end{equation}%
where the Fermi velocity is $\mathbf{v=\bigtriangledown }\epsilon $ and the
element of length along the FS is calculated numerically as $dp_{t}=\sqrt{%
\left( v_{y}/v_{x}\right) ^{2}+1}dp_{y}$. Finally one obtains the usual BCS
- like expression for $T_{c}$:

\begin{equation}
T_{c}=\frac{2\gamma _{E}}{\pi }\Omega \exp \left[ -\frac{1}{\lambda ^{d}}%
\right] \text{.}  \label{BCS}
\end{equation}%
The values of $T_{c}^{\max }$ at optimal doping are given as green lines in
Fig.5. This agrees well with numerical solution of the gap equation except
in the vicinity of the Lifshitz transition point at $t_{L}^{\prime }=-0.19$
where the naive BCS $T_{c}$ erroneously enters the strong coupling regime.
The reason of the breakdown is that the steepest descent approximation used
is invalidated near the $\mathbf{X}$ point due to the fact that the DOS
changes faster than $\tanh \left[ \varepsilon \right] /\varepsilon $ . In
particular this leads to the logarithmically divergent density of states.
Here a more sophisticated approximate calculation is required.

\begin{figure}[h]
\centering \includegraphics[width=12cm]{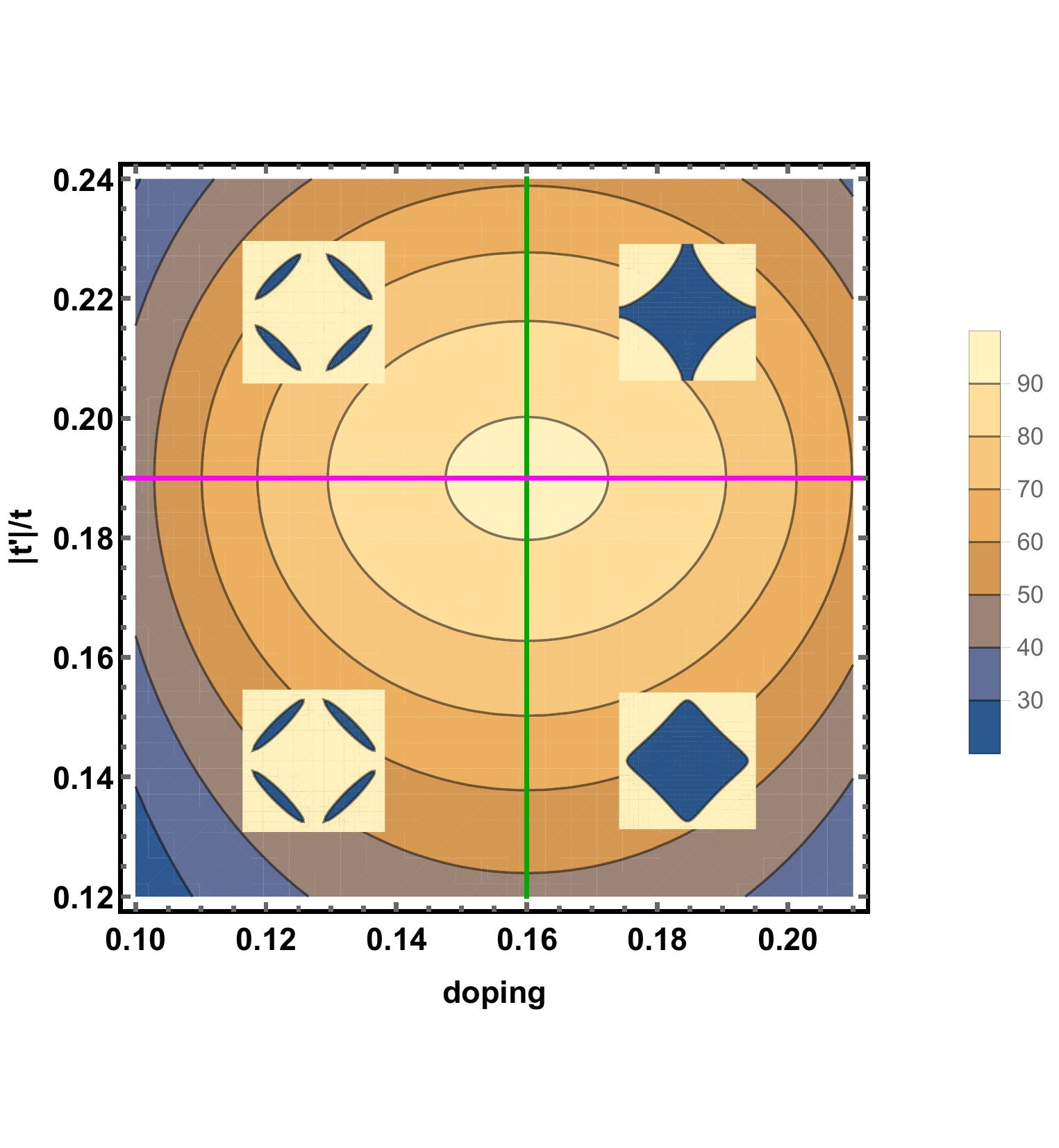}
\caption{The doping - hopping ($t^{\prime }/t$) phase diagram of the single
later cuprates. Materials with very high $T_{c}$ (Hg and Tl based cuprates)
are described by tp slightly above the critical open - close topological
transition hoping value of $t\prime /t=-0.19$. Lower $T_{c}$ La and Bi based
cuprates have much lower values around $t^{\prime }=0.12$ significantly
below the transition line (magenta). Near the optimal doping $p=0.16$ all
the materials with different hoping ratios undergo the second topological
transition (green) at which the open or closed Fermi surface fractures into
four small pockets. }
\end{figure}

\subsection{The van Hove singularity dominated critical system}

In a particular case of the close to open transition at $t^{\prime }=-0.19$,
the FS approaches the van Hove singularity at $\mathbf{k}=(0,\pi )$ and the
steepest descent approximation is invalid. However precisely in this case
one can simplify the gap equation by expanding the dispersion relation
around the dominant $\mathbf{X}$\textbf{\ }and\textbf{\ }$\mathbf{Y}$
points. Replacing the sums by integrals in Eq.(\ref{Tcgap}) and changing the
variables to $\left \{ \varepsilon =\epsilon \left( p_{x},p_{y}\right)
,p_{t}\right \} $, one obtains: 
\begin{equation}
1=-\frac{1}{\left( 2\pi \right) ^{2}\Omega ^{2}}\int_{\varepsilon =-\Omega
}^{\Omega }d\varepsilon \  \oint_{\varepsilon }dp_{t}\frac{1}{v\left(
p_{t}\right) }\rho _{\mathbf{p}}^{d}\frac{\tanh \left[ \varepsilon /2T_{c}%
\right] }{2\varepsilon }\text{.}  \label{gapvH}
\end{equation}%
This can be written via DOS, approximating $\rho _{\mathbf{p}}^{d}=\rho _{%
\mathbf{X}}^{d}=0.16\cdot \gamma $, see Fig. 4b, as 
\begin{equation*}
1=-\frac{\rho _{\mathbf{X}}^{d}}{4\Omega ^{2}}\int_{\varepsilon =-\Omega
}^{\Omega }d\varepsilon \frac{\tanh \left[ \varepsilon /2T_{c}\right] }{%
\varepsilon }D\left( \varepsilon \right) \text{.}
\end{equation*}%
Near the $\mathbf{X}$ point the DOS is:

\begin{eqnarray}
D\left( \varepsilon \right) &=&\frac{4}{\left( 2\pi \right) ^{2}\left(
1+2t^{\prime }/t\right) ^{3/2}}\ln \frac{b}{\varepsilon };  \label{DOS} \\
b &=&\left( 2\pi \right) ^{2}\left( 1-2t^{\prime }/t\right) ^{2}/\left( 1+%
\sqrt{2}\right) ^{2}\text{.}  \notag
\end{eqnarray}%
The critical temperature at the Lifshitz point ($t^{\prime }=t_{L}^{\prime }$%
, $\mu _{r}=\epsilon \left( \mathbf{X}\right) $, using methods described in
ref.\cite{Tsuei}, is%
\begin{equation}
T_{c}^{L}=t\exp \left[ -\sqrt{\log ^{2}\frac{2t}{\Omega }+\frac{2}{\lambda
_{L}}}\right] ,  \label{TcvH}
\end{equation}%
where the d-wave electron - phonon coupling strength is%
\begin{equation}
\lambda _{L}=\frac{\left \vert \rho _{\mathbf{X}}^{d}\right \vert }{\left(
2\pi \right) ^{2}\left( 1+2t_{L}^{\prime }/t\right) ^{3/2}\Omega ^{2}}\text{.%
}  \notag
\end{equation}

In the present case, $\Omega =50$ $meV$ and $t=460$ $meV$, one obtains $%
T_{c}^{\max }=82K$ (dashed light blue line in Fig. 5) compared to the
numerical result $T_{c}^{\max }=92K$. It might be thought of as an upper
cutoff on the naive d - wave BCS estimate in the topological transition
region. Just slightly away from it, $\left \vert t^{\prime }-t_{L}^{\prime
}\right \vert /t>0.1$, the BCS formula works well with values of $\lambda
^{d}=0.4-0.55$.

\section{Discussion and conclusions}

To conclude, large disparity of critical temperatures of one layered
cuprates is explained in the framework of the apical lateral longitudinal
phonon (ALLP) exchange d - wave superconductivity theory. To demonstrate the
basic principles we limited ourselves in this paper to a simple sufficiently
generic model of the electron gas in the $CuO$ planes: the fourfold
symmetric $t-t^{\prime }$ single band Hubbard model with on site repulsion
energy $U$ of moderate strength. It is shown that the highest $T_{c}^{\max } 
$ materials $HgBa_{2}CuO_{4+\delta }$ and $Tl_{2}Ba_{2}CuO_{6+\delta }$ ($%
>90K$) with a larger next to nearest neighbors hopping amplitude $t^{\prime
}/i\simeq 0.20$ are just above the topological transition from an open to a
close Fermi surface, while $La_{2-x}Sr_{x}CuO_{4+\delta }$ and ($<40K$) are
well below it at $t^{\prime }/t\simeq 0.12$ consistent with some first
principle calculations.

As was argued in previous papers\cite{I,preprint} the values of the
effective Coulomb repulsion $U/t$ away from the Mott insulator phase should
be in the intermediate coupling range, $1.5<U/t<3.5$, smaller than the
values for the parent materials. Only in this case the pseudogap and the
strange metal $d\rho /dT$ have experimentally observed order of magnitude.
Here we find that $U$ decreases monotonically with $t^{\prime }$, see Fig.
3b. The correlation between $t^{\prime }$ and $T_{c}^{\max }$ has been
noticed\cite{Pavarini01} on basis of early first principle calculations. Of
course other relevant for the APPL theory parameters of the one layer
cuprates, the lattice spacing $a$ and the distance of the apical oxygen $%
d_{A}$\ from the CuO plane and the phonon frequency $\Omega $ and the apical
oxygen charge $Z$ and nearest neighbor hoping amplitude $t$ are different.
Differences in $a$, $d_{A}$ and $t$ are known to be very small
experimentally. Less is known about the APPL electron - phonon coupling In
ref.\cite{I} we have estimated generally the range of $\Omega $ and $Z$
(determining the matrix element of the APPL and the CuO electron states) on
the basis of a simple microscopic Born-Meyer model described in Section II.
Looking at Fig.2, one notices that the Born - Meyer parameters for the
apical oxygen later of the two materials are almost identical. The parameter 
$A$ in Eq.(\ref{interatomic}) for La and Ba atoms are\cite{Abrahamson} $%
A_{La}=37.283eV$, $b_{La}=3.512$ $A^{-1}$ and $A_{La}=36.363$ $eV$, $%
b_{La}=3.514A^{-1}$ respectively. As a result both $\Omega $ and $Z$ are
practically the same.

Fig. 6 summarizes the APPL theory critical temperature dependence on both
doping and $t^{\prime }$. The topological phase diagram contains two \textit{%
different} Lifshitz transitions: at the critical doping $p_{opt}$ (the green
line) the FS fractures into four "pockets", while at critical next to
nearest neighbor hoping $t^{\prime }=t_{L}^{\prime }$ the center of FS moves
from $\Gamma $ to the $M$ point of Brillouin zone. Let us discuss some of
the limitations of the present model.

One is the role of the "unconventional" spin fluctuation d - wave pairing%
\cite{Kozik} which at intermediate coupling might still contribute and
"boost"\cite{I} $T_{c}$. In addition the restriction of the description of
the dispersion relation to the one band Hubbard model with just two
parameters $t,t^{\prime }$ for nearest neighbor and next to nearest neighbor
hopping obviously makes the model less realistic to quantitatively describe
real materials. If one considers adding more distant hopping terms like $%
t^{\prime \prime }$, the condition for the topological open to close Fermi
surface transitions, Eq.(\ref{vHcond}), is modified to $\mu _{r}=\epsilon
\left( \left \{ 0,\pi \right \} \right) =4t^{\prime }-4t^{\prime \prime }$.
It turns out that the $t^{\prime }\rightarrow t^{\prime }-t^{\prime \prime }$
substitution effectively describes an effect of this more complicated
dispersion relation (generally $t^{\prime }$ is negative, while $t^{\prime
\prime }$ is positive and smaller). Of course the situation in multi layered
cuprates is more involved due to inter - layer tunneling effects\cite%
{Lifshitz}.

The general conclusion is the following: critical temperature of transition
to superconductivity is always peaked at a topologically nontrivial
(Lifshitz) restructuring of the Fermi surface. When two such transitions
lines cross, see Fig. 6, the global maximum is achieved.

\textit{Acknowledgements. }

We are grateful Prof. D. Li, H.C. Kao, L. L.Wang, C. Q. Jin, J.Y. Lin and Y.
Yeshurun for helpful discussions. Work of B.R. was supported by NSC of
R.O.C. Grants No. 101-2112-M-009-014-MY3.\nolinebreak

\end{document}